\documentclass[pre]{revtex4}
\usepackage{amsmath}
\usepackage{graphicx}
\begin{document}
\title{One- and Two-dimensional Solitary Wave States in the Nonlinear Kramers Equation with Movement Direction as a Variable}
\author{Hidetsugu Sakaguchi and Kazuya Ishibashi}
\affiliation{Department of Applied Science for Electronics and Materials,
Interdisciplinary Graduate School of Engineering Sciences, Kyushu
University, Kasuga, Fukuoka 816-8580, Japan}
\begin{abstract}
We study self-propelled particles by direct numerical simulation of the nonlinear Kramers equation for self-propelled particles. In our previous paper, we studied self-propelled particles with velocity variables in one dimension. 
In this paper, we consider another model in which each particle exhibits directional motion. The movement direction is expressed with a variable $\phi$. We show that one-dimensional solitary wave states appear in direct numerical simulations of the nonlinear Kramers equation in one- and two-dimensional systems, which is a generalization of our previous result.  Furthermore, we find two-dimensionally localized states in the case that each self-propelled particle exhibits rotational motion. The center of mass of the two-dimensionally localized state exhibits circular motion, which implies collective rotating motion.  Finally, we consider a simple one-dimensional model equation to qualitatively understand the formation of the solitary wave state.    
\end{abstract}
\maketitle
\section{Introduction and the Nonlinear Kramers Equation}
The collective motion of self-propelled particles such as schools of fish and flocks of birds has been intensively studied since Vicsek and co-workers proposed a simple agent-based model for a large population of self-propelled particles~\cite{Vicsek,Vicsek2, Marchetti}.  Collective directional motion appears as a kind of order-disorder transition in Vicsek-type models~\cite{gre}.  In the disordered state, the directions of self-propelled particles are random. In the ordered state, a certain average direction appears in a large population of self-propelled particles. The spatial distribution is uniform in the ordered state. There is another nonuniform state called a solitary wave state. In the solitary wave state, localized regions of high density propagate similarly to a one-dimensional solitary wave~\cite{Chate,Bertin,Mishra,Gopinath, Bricard, Ihle}.  The solitary wave state was first found in direct numerical simulations based on the Vicsek model. 
In a previous paper, we showed that the solitary wave state appears in the one-dimensional nonlinear Kramers equation~\cite{Sakaguchi}. The nonlinear Kramers equation is a time evolution equation of the probability distribution for the position and velocity of self-propelled particles. In the previous model, we considered the probability distribution of the velocity $v_x$ and  position $x$.  However, there are three important variables, the direction of the velocity and the $x$, $y$ coordinates, in the original two-dimensional Vicsek model. The magnitude of the velocity is fixed to be a constant. The momentum direction is expressed by the angle $\phi$ from the $x$-axis. 
In this paper, we study the nonlinear Kramers equation for the angle $\phi$ and the two coordinates $x, y$. 

The model equations for elemental particles are expressed by the Langevin equation
\begin{subequations}
\begin{align}
\frac{dx_i}{dt}&=\cos\phi_i,\\
\frac{dy_i}{dt}&=\sin\phi_i,\\
\frac{d\phi_i}{dt}&=g\sum_{j=1}^Ne^{-\alpha\{2-\cos(2\pi(x_j-x_i)/L_x)-\cos(2\pi(y_j-y_i)/L_y)\}}\sin(\phi_j-\phi_i)+\xi_i(t), 
\end{align}
\end{subequations}
where $x_i$, $y_i$, and $\phi_i$ are the $x$, $y$ coordinates and the angle of the movement of the $i$th element, respectively, $L_x\times L_y$ denotes the system size, and $\xi_i(t)$ is Gaussian white noise satisfying $\langle \xi_i(t)\xi_j(t')\rangle=2T\delta_{i,j}\delta(t-t')$. 
The magitude of the velocity vector is fixed to 1. 
At $g=0$, each elemental particle moves independently in the direction of $\phi_i$ with velocity 1. The nonlocal interaction in the spatially periodic two-dimensional system of size $L_x\times L_y$ is expressed by the first term in the summation on the right-hand side of Eq.~(1c).  For $g>0$, the direction of motion tends to be mutually aligned. The last noise term in Eq.~(1c) makes the direction of motion random. A phase transition is expected to occur as a result of the competition of the two effects.   Equation (1c) is very similar to the nonlocally coupled Kuramoto model in that the mutual interaction is expressed as a sinusoidal function of the phase difference~\cite{Kuramoto, Kuramoto2,Strogatz}. 

The Kramers equation corresponding to the Langevin equation is expressed as
\begin{eqnarray}
& &\frac{\partial P}{\partial t}=-\frac{\partial}{\partial x}\left (\cos\phi P\right )-\frac{\partial}{\partial y}\left (\sin\phi P\right )+T\frac{\partial^2P}{\partial \phi^2}\nonumber\\
& &-\frac{\partial}{\partial \phi}\left  [\left \{g\int_0^{L_x}\int_0^{L_y}e^{-\alpha\{2-\cos(2\pi (x^{\prime}-x)/L_x)-\cos(2\pi (y^{\prime}-y)/L_y)\}}r(x^{\prime},y^{\prime})\sin(\bar{\phi}(x^{\prime},y^{\prime}) -\phi)dx^{\prime}dy^{\prime}\right \}P\right ],\nonumber\\ 
\label{kr}
\end{eqnarray} 
where $P(x,y,\phi,t)$ is the probability density function and $r(x,y)e^{i\bar{\phi}(x,y)}=\int_{0}^{2\pi}P(x,y,\phi)e^{i\phi}d\phi$. 
Here, we have assumed a kind of mean-field approximation in that the summation is replaced by the integral using the density and the average direction $\bar{\phi}$. This is an approximation in that some fluctuation effects and correlation effects between the direction of motion and the density are neglected.
The Kramers equation is a nonlinear equation because $r(x,y)e^{i\bar{\phi}}$ in the fourth term of the right-hand side of Eq.~(2) is expressed with the average of $e^{i\phi}$ with respect to $P(x,y,\phi, t)$. 
Since the nonlinear Kramers equation is a deterministic equation, the phase transitions can be treated as bifurcations in the nonlinear equation. 
In this paper, we assume the normalization condition $\int_0^{L_x}\int_0^{L_y}\int_{0}^{2\pi}P(x,y,\phi)d\phi dxdy=1$. 
The integral kernel $e^{-\alpha\{2-\cos(2\pi (x^{\prime}-x)/L_x)-\cos(2\pi(y^{\prime}-y)\}}$ can be approximated as the Gaussian function $e^{-(2\pi^2\alpha/L_x^2)(x^{\prime}-x)^2-(2\pi^2\alpha/L_y^2)(y^{\prime}-y)^2}$ if $x^{\prime}-x$ and $y^{\prime}-y$ are sufficiently small. The Gaussian kernel is used in the numerical simulations of two-dimensional systems. 

\section{One-Dimensional System}
Firstly, we consider the one-dimensional system 
\begin{equation}
\frac{\partial P}{\partial t}=-\frac{\partial}{\partial x}\left (\cos\phi P\right )-\frac{\partial}{\partial \phi}\left  [\left \{g\int_0^{L_x}e^{-\alpha\{1-\cos(2\pi (x^{\prime}-x)/L_x)\}}r(x^{\prime})\sin(\bar{\phi}(x^{\prime}) -\phi)dx^{\prime}\right \}P\right ]+T\frac{\partial^2P}{\partial \phi^2} \label{kr2},
\end{equation} 
where $r(x)e^{i\bar{\phi}(x)}=\int_{0}^{2\pi}P(x,\phi)e^{i\phi}d\phi$.
 Equation (3) can be derived from Eq.~(2) if $P(x,y,\phi)$ is independent of $y$, or the distribution is uniform in the $y$-direction and $g\int_0^{L_y}\exp[-\alpha\{1-\cos(2\pi(y^{\prime}-y)/L_y)\}]dy^{\prime}$ is set to $g$. 
The velocity $v_x$ of each self-propelled particle is expressed as $\cos\phi$.  The normalization condition is $\int_0^{L_x}\int_{0}^{2\pi}P(x,\phi)d\phi dx=1$. There is a uniform solution: $P(x,\phi)=1/(2\pi L_x)$. In this uniform state, the average velocity $\langle \cos\phi \rangle$ is 0. 
However, for sufficiently large $g$, the uniform state becomes unstable and the average velocity  $\langle \cos\phi \rangle$ becomes nonzero.  

If a stationary solution $P_0(x,\phi)$ does not depend on $x$, $P_0(\phi)$ is expressed by the thermal equilibrium distribution as
\begin{equation}
P_0(\phi)\propto e^{K\langle \cos\phi\rangle \cos\phi/T},
\end{equation}
where $K=(g/L_x)\int_0^{L_x}e^{-\alpha\{1-\cos(2\pi x/L_x)\}}dx$. 

Because $\langle\cos\phi\rangle=\int_0^{2\pi}P_0(\phi)\cos\phi d\phi/\int_0^{2\pi}P_0(\phi)d\phi$,   
$\langle\cos\phi\rangle$ is expressed as
\begin{equation}
\langle\cos\phi\rangle=\frac{\int_0^{2\pi}e^{K\langle\cos\phi\rangle\cos\phi/T}\cos\phi d\phi}{\int_0^{2\pi}e^{K\langle\cos\phi\rangle\cos\phi/T}d\phi}.
\end{equation}
This is a self-consistent equation for $\langle\cos\phi\rangle$. 
If $\langle\cos\phi\rangle$ is sufficiently small, the self-consistent equation can be approximated as 
\begin{equation}
\langle\cos\phi\rangle\simeq\frac{\int_0^{2\pi}K\langle\cos\phi\rangle\cos^2\phi d\phi/T}{2\pi}=\frac{K}{2T}\langle\cos\phi\rangle+O(\langle\cos\phi\rangle)^3).\end{equation}
The nonzero $\langle\cos\phi\rangle$ appears for $K>K_c=2T$. That is, the critical value of $g$ is 
\begin{equation}
g=\frac{2TL_x}{\int_0^{L_x}e^{-\alpha\{1-\cos(2\pi x/L_x)\}}dx}=\frac{4\pi T}{\int_0^{2\pi}e^{-\alpha(1-\cos\phi)}d\phi}.
\end{equation} 
The critical line does not depend on the system size $L_x$. 
\begin{figure}[t]
\begin{center}
\includegraphics[height=5.cm]{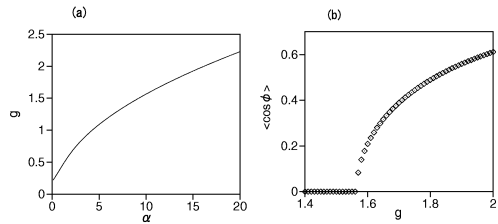}
\end{center}
\caption{(a) Critical line in the parameter space of $(\alpha,g)$ at $T=0.1$. (b) Average velocity $\langle \cos\phi\rangle$ as a function of $g$ at $T=0.1$.}
\label{f1}
\end{figure}
\begin{figure}
\begin{center}
\includegraphics[height=5.cm]{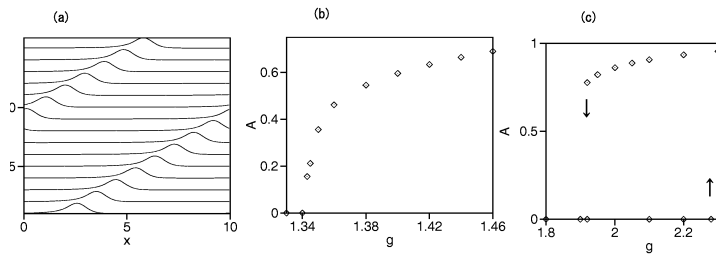}
\end{center}
\caption{(a) Fourier amplitude $A$ for the local order parameter $\langle \cos\phi(x)\rangle$ as a function of $g$ at $\alpha=5$, $T=0.1$, and $L_x=10$. (b) Fourier amplitude $A$ for the local order parameter $\langle \cos\phi(x)\rangle$ as a function of $g$ at $\alpha=15$, $T=0.1$, and $L_x=10$.  (c) Phase diagram in the parameter space of $(\alpha,g)$. }
\label{f2}
\end{figure}
\begin{figure}
\begin{center}
\includegraphics[height=4.5cm]{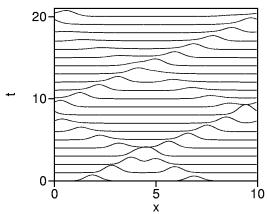}
\end{center}
\caption{Head-on collision of two solitary waves at $g=2$, $\alpha=5$, $T=0.1$, and $L_x=10$. }
\label{f3}
\end{figure}

Figure 1(a) shows the critical line of Eq.~(7) in the parameter space of $(\alpha,g)$ at $T=0.1$. A nonzero $\langle\cos\phi\rangle$ appears above the critical line. 
Figure 1(b) shows the average velocity $\langle \cos\phi\rangle$ as a function of $g$ for $L_x=10$ and $\alpha=10$. The average velocity increases continuously from 0.  

For larger $g$, the spatially uniform state can be unstable and a solitary wave state appears.   
The spatial inhomogeneity can be evaluated by the Fourier amplitude of the local order parameter
\begin{equation}
A=\left|\frac{1}{L_x}\int_0^{L_x}\langle \cos\phi(x)\rangle e^{i2\pi x/L_x}dx\right |,
\end{equation}
where $\langle \cos(\phi(x))\rangle=\int_0^{2\pi}P(x,\phi)\cos\phi d\phi/\int_0^{2\pi}P(x,\phi)d\phi$. 
Figure 2(a) shows the Fourier amplitude $A$ as a function of $g$ for $L_x=10$, $\alpha=5$, and $T=0.1$. The Fourier amplitude $A$ increases from 0 continuously at $g\simeq 1.34$. Figure 2(b) shows the Fourier amplitude $A$ as a function of $g$ for $\alpha=15$, $L_x=10$, and $T=0.1$. At these parameter values, $A$ jumps from 0 to 0.95 at $g=2.29$ when $g$ increases, and $A$ jumps from 0.78 to 0 at $g=1.91$.  The transitions are discontinuous and hysteresis occurs. 

Figure 2(c) shows the phase diagram for $L_x=10$ and $T=0.1$.  There are three states: a disordered state `D', a spatially uniform ordered state `O', and a solitary wave state `S'. For $\alpha<7.5$, the transitions from the disordered state to the spatially uniform ordered state and from the spatially uniform  ordered state to the solitary wave state are continuous. On the other hand, for $\alpha>7.5$, the transition from the disordered state to the spatially uniform ordered state is continuous; however, the transition from the spatially uniform ordered state to the solitary wave state is discontinuous, which is denoted by the dashed line. The transition from the solitary wave state to the spatially uniform ordered state or the disordered state is denoted by the dotted line in Fig.~2(d). In the previous paper, we constructed a similar phase diagram for the nonlinear Kramers equation with velocity as the variable~\cite{Sakaguchi}. 

Figure 3 shows a head-on collision of two solitary waves with slightly different amplitudes at $g=2$, $L_x=10$, $\alpha=5$, and $T=0.1$. The two solitary waves interpenetrate each other at the first collision. However, the amplitude difference increases at successive collisions and only one solitary wave survives after a long time. This behavior is slightly different from the head-on collision of two solitary waves in our previous model~\cite{Sakaguchi}, where merging occurred at the first collision. The reason for the difference is not clear; however, it is not so surprising because various phenomena such as pair annihilation, interpenetration, and the formation of a bound state occur at the head-on collision of two general dissipative solitons depending on the control parameters~\cite{Krischer, Descalzi}. This type of behavior is similar to the numerical result obtained with a model based on the kinetic theory for the Vicsek model by Ihle~\cite{Ihle}. 
\section{Two-Dimensional System}
 Direct numerical simulation of two-dimensional nonlinear Kramers equations is possible; however, it takes a very long time because the double integration in Eq.~(2) is necessary at each point $(x,y)$. In this paper, a two-dimensional system of size $L\times L$  is discretized with $\Delta x=L/N$, and  the double integration is approximated by a local summation for neighboring  sites. That is, the double integration in Eq.~(2) is replaced with
\[
\sum_{(i^{\prime},j^{\prime})\in N(i,j)} e^{-\alpha^{\prime} \Delta x^2\{(i^{\prime}-i)^2+(j^{\prime}-j)^2\}}r(i^{\prime},j^{\prime})\sin(\bar{\phi}(i^{\prime},j^{\prime}) -\phi),
\] 
where $i^{\prime}=x^{\prime}/\Delta x, j^{\prime}=y^{\prime}/\Delta x, i=x/\Delta x, j=y/\Delta x$, and  $r(x^{\prime},y^{\prime})$ and $\bar{\phi}(x^{\prime},y^{\prime})$ are respectively expressed as $r(i^{\prime},j^{\prime})$ and $\bar{\phi}(i^{\prime},j^{\prime})$. The summation is taken for  61 neighboring sites satisfying $(i^{\prime}-i)^2+(j^{\prime}-j)^2\le 18$ around each $(i,j)$ site. The partial derivative $\partial P/\partial x $ is calculated with the central difference $(P(i+1,j)-P(i-1,j))/(2\Delta x)$. 
The parameter $\alpha^{\prime}$ is approximately expressed as $(2\pi^2/L^2)\alpha$ using $\alpha$ in Eq.~(2). Similarly, the angle variable $\phi$ is discretized with $\Delta\phi=2\pi/M$. The following numerical simulation is carried out with $N\times N\times M=99\times 99\times 99$. 
Periodic boundary conditions are imposed for $x$, $y$, and $\phi$.  

Figure 4(a) shows a three-dimensional plot of the density $\rho(x,y)$ at $\alpha^{\prime}=80$, $L=2.5$, $T=0.2$, and $g=10$. The density is localized in the $x$-direction and uniform in the $y$-direction. The localized pulse is propagating in the $x$-direction. This type of one-dimensional solitary wave state corresponds to the solitary wave state found in the numerical simulations based on the Vicsek model. There are some approximate theories for the solitary wave state~\cite{Caussin}; however, the mechanism for the formation of the solitary wave state is not clear. Although there is a report that the straight bands become unstable and chaotic patterns appear in numerical simulations for larger systems based on the Vicsek model, we did not observe the instability in our deterministic model. 
\begin{figure}[h]
\begin{center}
\includegraphics[height=5.5cm]{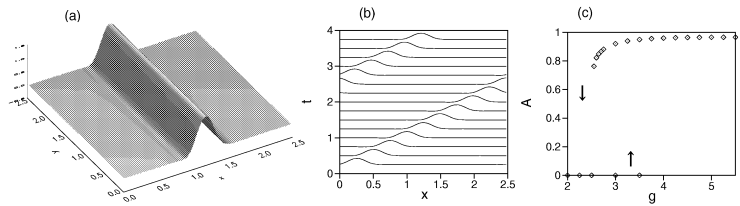}
\end{center}
\caption{(a) 3D plot of $\rho(x,y)$ at $\alpha^{\prime}=80$, $L=2.5$, $T=0.2$, and $g=10$.  (b) Fourier amplitude $A$ as a function of $g$ at $\alpha^{\prime}=80$, $L=2.5$, and $\Delta x=2.5/99$. }
\label{f4}
\end{figure}

Figure 4(b) shows the Fourier amplitude as a function of $g$ at $\alpha^{\prime}=80$, $L=2.5$, and $T=0.2$.
The spatially uniform ordered state becomes unstable at $g=3.6$. 
The solitary wave state jumps to the disordered state at $g=2.5$. There is hysteresis between $g=2.6$ and 3.5. 
\section{Two-Dimensionally Localized Solitary Wave State for Self-Rotating Particles}
In the Vicsek model, each self-propelled particle moves in a certain direction. However, another class of self-propelled particles that change their direction autonomously is also interesting. Such active matter is called chiral active matter. For example, the circle swimmers E. coli~\cite{Berg} and magnetotactic bacteria in rotating external fields~\cite{Erglis} are included in chiral active matter.  The collective motion of chiral active matter was studied by several authors~\cite{Yeo}. Liebchen and Levis found a macrodroplet and microflock pattern in a Vicsek type model for self-rotating particles~\cite{Liebchen}.  
In the macrodroplet state, self-rotating particles gather and make a circular cluster, which is a two-dimensionally localized state. The circular cluster rotates coherently. Here we study the two-dimensionally localized solitary wave state using the nonlinear Kramers equation. 

We consider self-rotating particles that obey the following Langevin equation instead of Eq.~(1): 
\begin{eqnarray}
\frac{dx_i}{dt}&=&\cos\phi_i,\nonumber\\
\frac{dy_i}{dt}&=&\sin\phi_i,\nonumber\\
\frac{d\phi_i}{dt}&=&\omega+g\sum_{j=1}^Ne^{-\alpha\{2-\cos(2\pi(x_j-x_i)/L_x)-\cos(2\pi(y_j-y_i)/L_y)\}}\sin(\phi_j-\phi_i)+\xi_i(t).
\end{eqnarray}
The direction of motion changes with time spontaneously, whose natural frequency is $\omega$. Equation (9) is reduced to Eq.~(1) if $\omega=0$.  
This equation is similar to the model equation of Liebchen and Levis. 
Sumino et al. studied another type of model of self-rotating particles that form a vortex lattice~\cite{Sumino}.   
The nonlinear Kramers equation corresponding to Eq.~(9) is expressed as
\begin{eqnarray}
& &\frac{\partial P}{\partial t}=-\frac{\partial}{\partial x}\left (\cos\phi P\right )-\frac{\partial}{\partial y}\left (\sin\phi P\right )+T\frac{\partial^2P}{\partial \phi^2}\nonumber\\
& &-\frac{\partial}{\partial \phi}\left  [\left \{\omega+g\int_0^{L_x}\int_0^{L_y}e^{-\alpha\{2-\cos(2\pi (x^{\prime}-x)/L_x)-\cos(2\pi (y^{\prime}-y)/L_y)\}}r(x^{\prime},y^{\prime})\sin(\bar{\phi}(x^{\prime},y^{\prime}) -\phi)dx^{\prime}dy^{\prime}\right \}P\right ]. \nonumber\\
\label{kr2}
\end{eqnarray}

\begin{figure}[h]
\begin{center}
\includegraphics[height=6.5cm]{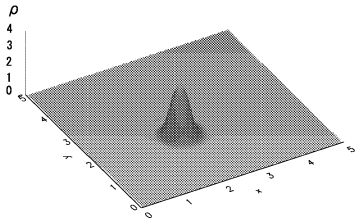}
\end{center}
\caption{3D plots of $\rho(x,y)$ for $\alpha^{\prime}=80$, $L=5$, $g=15$, $\omega=0.2$, and $T=0.2$. }
\label{f51}
\end{figure}
\begin{figure}[h]
\begin{center}
\includegraphics[height=6.5cm]{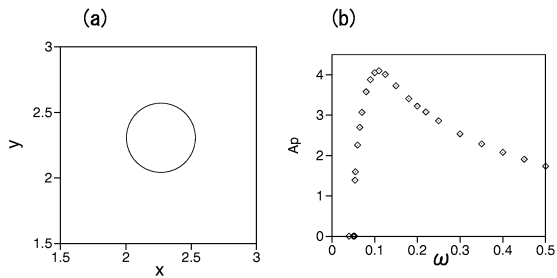}
\end{center}
\caption{ (a) Trajectory of the center of mass of the two-dimensionally localized state. (b) Peak amplitude $A_p$ of $\rho(x,y)$ as a function of $\omega$ for $\alpha^{\prime}=80$, $L=5$, $g=15$, and $T=0.2$. }
\label{f52}
\end{figure}
Figure 5 shows 3D plots of the density $\rho(x,y)$ for $\alpha^{\prime}=80$, $L=5$, $g=15$, and $T=0.2$. 
The density and the local order parameter are localized in both the $x$ and $y$ directions. Figure 6(a) shows the trajectory of the center of mass defined by
\begin{equation}
(X,Y)=\left (\int_0^L\int_0^L\rho(x,y)xdxdy/\int_0^L\int_0^L\rho(x,y)dxdy,\int_0^L\int_0^L\rho(x,y)ydxdy/\int_0^L\int_0^L\rho(x,y)dxdy\right ).
\end{equation}
The solitary wave state is rotating around $(X_0,Y_0)$, where $X_0=Y_0=2.27$. The radius of rotation is 0.265 and the frequency of the rotation is 3.0. That is, a spatially localized and collectively rotating state appears. In this state, self-rotating particles make a flock and exhibit synchronous rotation. The two-dimensionally localized solitary wave state is closely related to the traveling band state at $\omega=0$; however, the straight traveling band state cannot survive for $\omega>0$, because the direction of motion of each self-propelled particle changes with time.  This solitary wave state corresponds to the macrodroplet state in the Vicsek type model of Liebchen and Levis. We observed some states in which there were several spots corresponding to the microflock pattern; however, we have not studied the states including several spots in detail, which is left to future study.  Figure 6(b) shows the peak amplitude of the density $\rho(x,y)$ as a function of $\omega$ for $\alpha^{\prime}=80$, $L=5$, $g=15$, and $T=0.2$. 
There is a peak around $\omega=0.11$. The two-dimensional solitary wave state disappears at $\omega=0.053$. For $\omega\le 0.053$, a spatially uniform ordered state appears, where the average movement direction $(\langle \cos\phi\rangle, \langle \sin\phi\rangle)$ changes with time. The transition to the spatially uniform ordered state is discontinuous. The spatially uniform ordered state becomes unstable and jumps to the two-dimensional solitary wave state at $\omega=0.086$ when $\omega$ gradually increases from 0.05. 
\section{Simple Model for Solitary Wave State} 
The mechanism of the instability from the ordered state to the solitary wave state is still not clear. In this section, we study a simple one-dimensional model equation to understand the instability qualitatively. We assume that each self-propelled particle takes one of two velocities, $v_0$ or $-v_0$, and the velocity of each particle changes from $v_0$ to $-v_0$ with transition probability $r_{+}$ and from $-v_0$ to $v_0$ with  transition probability $r_{-}$. Then, the probability densities $P_{\pm}(x)$ of velocities $\pm v_0$ obey the model equation
\begin{eqnarray}
\frac{\partial P_{+}}{\partial t}&=&-v_0\frac{\partial P_{+}}{\partial x}+D\frac{\partial^2 P_{+}}{\partial x^2}+r_{-}P_{-}-r_{+}P_{+},\nonumber\\
\frac{\partial P_{-}}{\partial t}&=&v_0\frac{\partial P_{-}}{\partial x}+D\frac{\partial^2 P_{-}}{\partial x^2}+r_{+}P_{+}-r_{-}P_{-},
\end{eqnarray} 
where $D\partial^2 P_{\pm}/\partial x^2$ are artificial diffusion terms to suppress divergence.  Furthermore, we assume that $r_{-}=e^{g(P_{+}-P_{-})}$ and $r_{+}=e^{-g(P_{+}-P_{-})}$. This equation represents simple dynamics of the mean-field type Ising model if spatial uniformity is assumed. At the equilibrium state, the relation
\begin{equation}
\frac{P_{-}}{P_{+}}=\frac{r_{+}}{r_{-}}=e^{-2g(P_{+}-P_{-})}
\end{equation}
is satisfied. 
The spatially uniform solutions $P_{0+}$ and $P_{0-}$ are obtained from this self-consistent equation. 
If the normalization $\int_0^L(P_{+}(x)+P_{-}(x))dx=1$ is assumed,
the spatially uniform disordered state $P_{0+}=P_{0-}=1/(2L)$ becomes unstable at $g=L$, and the spatially uniform ordered state $P_{0+}\ne P_{0-}$ appears for $g>L$. 
From Eq.~(12), perturbations of the form $\delta P_{+}e^{ikx+\lambda t}$ and $\delta P_{-}e^{ikx+\lambda t}$ obey
\begin{eqnarray}
\lambda\delta P_{+}&=&(-ikv_0-Dk^2+a_{11})\delta P_{+}+a_{12}\delta P_{-},\nonumber\\\lambda\delta P_{-}&=&-a_{11}\delta P_{-}+(ikv_0-Dk^2-a_{12})\delta P_{-},
\end{eqnarray} 
where 
\[a_{11}=-e^{-g(P_{0+}-P_{0-})}+ge^{g(P_{0+}-P_{0-})}P_{0-}+ge^{-g(P_{0+}-P_{0-})}P_{0+},\]           
\[a_{12}=e^{g(P_{0+}-P_{0-})}-ge^{g(P_{0+}-P_{0-})}P_{0-}-ge^{-g(P_{0+}-P_{0-})}P_{0+}.\]           
The eigenvalue $\lambda(k)$ is expressed as 
\[
\lambda(k)=\frac{-(a_{12}-a_{11})\pm\sqrt{(a_{12}-a_{11})^2-4k^2v_0^2-4ikv_0(a_{11}+a_{12})}}{2}-Dk^2.
\]
The linear growth rate or the real part of $\lambda$ can be explicity written as\begin{equation}
{\rm Re}\lambda(k)=\frac{-(a_{12}-a_{11})+\beta}{2}-Dk^2,
\end{equation}
where
\begin{equation}
\beta=\left [\frac{(a_{12}-a_{11})^2-4k^2v_0^2+\sqrt{\{(a_{12}-a_{11})^2-4k^2v_0^2\}^2+16k^2v_0^2(a_{11}+a_{12})^2}}{2}\right ]^{1/2}.
\end{equation}
Figure 7(a) shows the relationship between $k$ and Re$\lambda(k)$ for $D=0$ and 0.01 at $g=8$, $L=5$, and $v_0=1$. The linear growth rate becomes positive, which implies that the spatially uniform state is unstable. At $k=0$, Re$\lambda$=0.  At $D=0$,  Re$\lambda\rightarrow a_{11}$  for $k\rightarrow \infty$, which can be shown using Eqs.~(15) and (16). 
The instability originates from the terms $\mp v_0\partial P_{\pm}/\partial x$ in Eq.~(12). Figure 7(b) shows the peak value of Re$\lambda(k)$ as a function of $g$ for $D=0.01$, $L=5$, and $v_0=1$.  The spatially uniform state is unstable for $5<g<12.7$. 
\begin{figure}[h]
\begin{center}
\includegraphics[height=5.5cm]{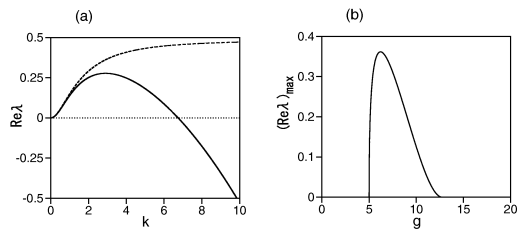}
\end{center}
\caption{(a) Re$\lambda(k)$ as a function $k$ for $D=0$ and 0.01 at $g=8$, $L=4$, and $v_0=1$. (b) Peak value of Re$\lambda(k)$ as a function $g$ for $D=0.01$, $L=4$, and $v_0=1$. }
\label{f6}
\end{figure}

Figure 8(a) shows the time evolution of the density $\rho(x)=P_{+}(x)+P_{-}(x)$ at $g=8$, $L=5$, $v_0=1$, and $D=0.01$. A propagating solitary wave state appears. The solitary wave state propagates with velocity 1.04, which is slightly larger than $v_0=1$.  At $D=0$, Eq.~(12) exhibits divergence and the steadily propagating solitary wave state cannot be obtained.  Figure 8(b) shows the Fourier amplitude $A=|\int_0^L(P_{+}-P_{-})e^{2\pi i x/L}dx|$ of the local order parameter $P_{+}-P_{-}$ as a function of $g$ at $L=5$, $v_0=1$, and $D=0.01$. The solitary wave state is stable for $2.3<g<16.1$.  The solitary wave state and the spatially uniform disordered state are bistable for $2.3<g<5$, and the solitary wave state and the spatially uniform ordered state are bistable for $12.7<g<16.1$. For $g>16.1$, only  the spatially uniform ordered state is stable. This might be related to the previous numerical results showing that the solitary wave states appear near the transition range between the disordered state and the ordered state.
\begin{figure}[h]
\begin{center}
\includegraphics[height=5.5cm]{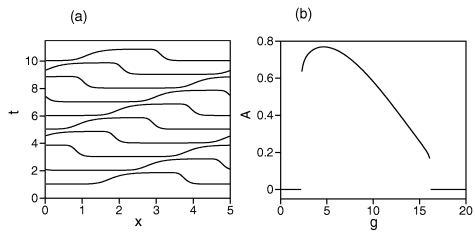}
\end{center}
\caption{ (a) Time evolution of the density $P_{+}+P_{-}$ at $g=8$, $L=5$, $v_0=1$ and $D=0.01$ for Eq.~(12). (b) Fourier amplitude $A$ of the order parameter $P_{+}(x)-P_{-}(x)$ as a function of $g$ at $L=5$, $v_0=1$, and $D=0.01$ for Eq.~(12). }
\label{f62}
\end{figure}

We consider that the instability of the spatially uniform ordered state in Eq.~(2) is also caused by the drift term $-\partial/\partial x( \cos\theta P)$, and that the nonlocal coupling term in Eq.~(2) might play a role of the artificial diffusion term in Eq.~(12) to suppress the divergence.  
\section{Summary}
We have studied the nonlinear Kramers equation with the movement direction as a variable in one and two dimensions. We have reproduced a one-dimensional solitary wave state, which appears from the instability of the spatially uniform ordered state.  This is a generalization of our previous result. Furthermore, we have found a two-dimensionally localized solitary wave state in the case that elemental particles are self-rotating. Such a two-dimensionally localized solitary wave state might be interesting for circulating living species such as E. coli. Finally, we have constructed a simple one-dimensional model equation and found that the instability of the spatially uniform ordered state is caused by the drift term.   

\end{document}